\documentclass[a4paper]{jpconf}
\usepackage{graphicx}
\usepackage{subfigure}

\begin{document}
\title{Refactoring, reengineering and evolution: paths to Geant4 uncertainty quantification and performance improvement}

\author{M Bati\v{c}$^{1,2}$, M~Begalli$^3$, M~Han$^4$, S~Hauf$^5$, G~Hoff$^{1,6}$, C~H~Kim$^4$, M~Kuster$^7$,
M~G~Pia$^1$, P~Saracco$^1$, H~Seo$^4$, G~Weidenspointner$^{8,9}$, A~Zoglauer$^{10}$}

\address{$^1$ INFN Sezione di Genova, Genova 16146, Italy}
\address{$^2$ Jozef Stefan Institute, 1000 Ljubljana, Slovenia}
\address{$^3$12UERJ, 20550-013, Rio de Janeiro, RJ, Brazil}
\address{$^4$  Hanyang University, Seoul 133-791, Korea}
\address{$^5$ Technische Universit\"at Darmstadt, IKP, Germany}
\address{$^6$ Pontificia Universidade Catolica do Rio Grande do Sul, Porto Alegre, Brazil.}
\address{$^7$ European XFEL GmbH, Hamburg, Germany}
\address{$^8$ Max-Planck-Institut f\"ur extraterrestrische Physik, 85740 Garching, Germany}
\address{$^9$ MPI Halbleiterlabor, 81739 M\"unchen, Germany}
\address{$^{10}$ Space Sciences Laboratory, University of California at Berkeley, Berkeley, CA 94720, USA}

\ead{Maria.Grazia.Pia@cern.ch}

\begin{abstract}
Ongoing investigations for the improvement of Geant4 accuracy and computational performance 
resulting by refactoring and reengineering parts of the code are discussed.
Issues in refactoring that are specific to the domain of physics simulation are 
identified and their impact is elucidated.
Preliminary quantitative results are reported.

\end{abstract} 

\section{Introduction}

The Geant4 \cite{g4nim,g4tns} simulation toolkit is nowadays a mature software
system: at the time of writing this paper, its reference publication \cite{g4nim} has
collected more than 3000 citations \cite{wos}.
The development of Geant4 started in 1994 as the RD44 \cite{rd44} project and its
first version was released at the end of 1998; since then, Geant4 has
been used in a wide variety of experimental applications, while further code
development continued, also motivated by new requirements originating 
from the experimental community.

Over the 18 years elapsed since the start of Geant4 development, the object
oriented paradigm has evolved from the status of pioneering technology into
established methods and software design techniques, while new compilers nowadays
support features of the C++ language that were not practically available in
earlier versions.

Methods and techniques have been developed over the years to cope effectively
with the evolution of large scale object oriented systems by providing
guidance for the improvement of the design of existing software; some of them
are now well established components of the software engineering body of
knowledge, and are documented in classical textbooks \cite{fowler},
\cite{feathers}, \cite{oorp}.

While the architectural design of Geant4 established in RD44 has demonstrated
its soundness by supporting the growth of the toolkit and its applications in
multidisciplinary environments, Geant4 could profit from exploiting established
refactoring and reengineering techniques to improve the design in some parts of
the code, especially those that have been subject to extensive evolution in
recent years, or that should accommodate new experimental requirements.

A project in progress investigates the benefits that could derive from the
exploitation of these techniques, namely in Geant4 physics domain: this
investigation is not limited to evaluating effects that are typically associated
with design improvements, such as ease of maintenance and facilitation of
further extensions of functionality, but it also estimates their impact on the
quantification of Geant4 simulation accuracy and its computational performance.
This project also explores issues, and methods to address them, that, while
conceptually similar to those encountered in conventional refactoring projects,
are specific to the environment of a large scale physics software system such as
Geant4.

This conference paper summarizes the main ideas underlying the ongoing R\&D
(research and development) pursued by the authors, and a few initial results of
the activity in progress; extensive details are meant to be documented in
dedicated publications in scholarly journals.

\section{Vision}

The activities documented in this paper - refactoring and reengineering parts of
Geant4 code - are carried out within the scope of a wider scientific research vision,
focused on the investigation of fundamental topics in particle transport.
Research is articulated over two main areas, that are logically and technically
intertwined: the assessment of the state-of-the-art in physics modeling for
particle transport (with consequent improvement of Geant4 physics to
reflect the state-of-the-art, if not yet achieved), and the objective quantification of
the uncertainty of simulation results.

Refactoring and reengineering techniques support this scientific vision by
contributing to improve the software design, hence the transparency of Geant4
physics modeling, the capability of quantifying its accuracy at a fine grained
level of detail, and the agility towards implementing state-of-the-art physics
in a computationally effective environment.

\begin{figure}
\centerline{\includegraphics[angle=0,width=10cm]{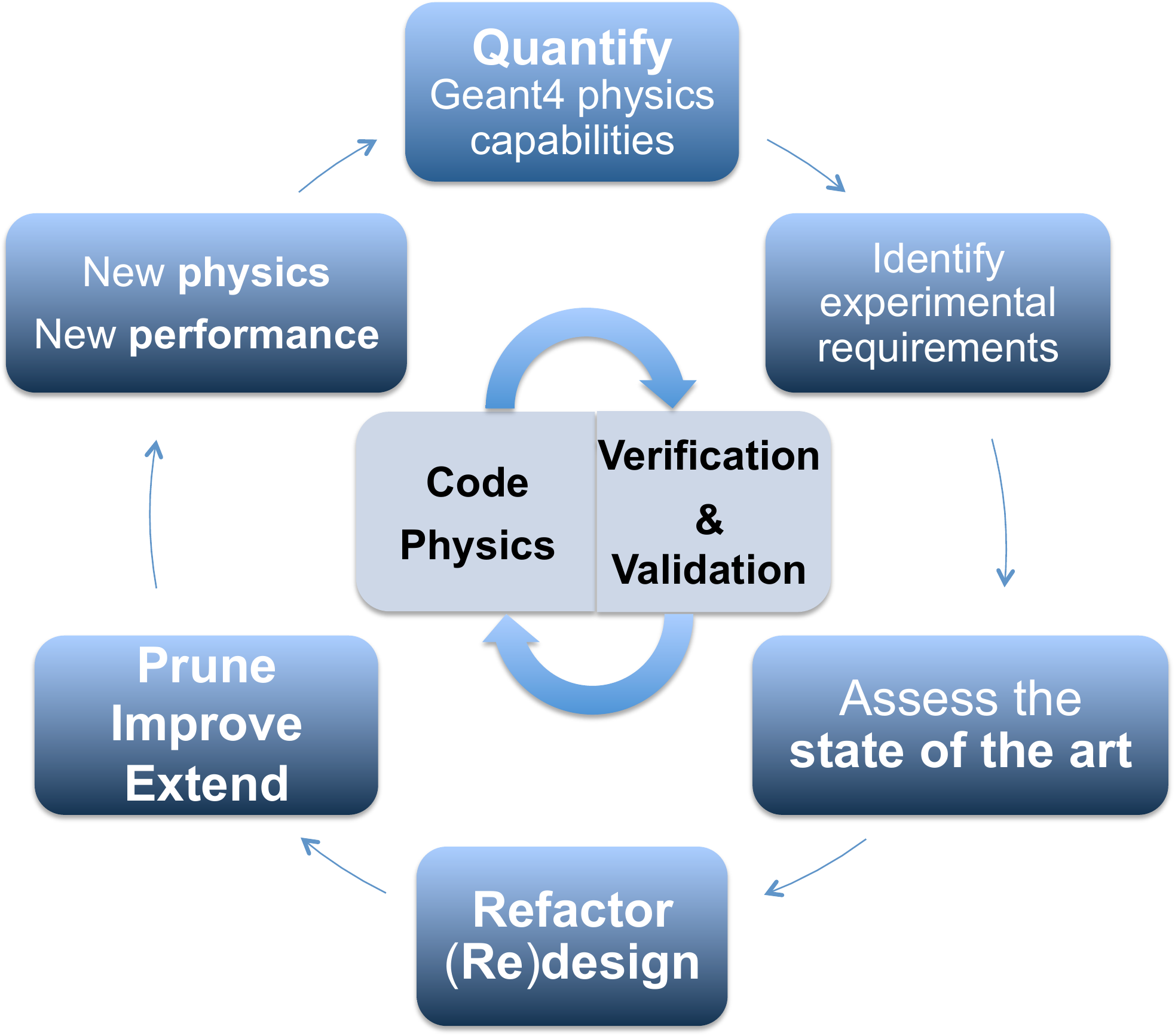}}
\caption{Simplified illustration of the life-cycle model adopted in the
investigation of the state-of-the-art in physics models for particle transport.}
\label{fig_lifecycle}
\end{figure}

The project adopts an iterative-incremental life-cycle model \cite{up}, which is
illustrated in figure \ref{fig_lifecycle}: while it is supported by a broad
scientific vision, concrete deliverables are produced in the course of the
activity, which are practically usable in the current simulation environment and
respond to existing experimental issues.

\section{Technical matters}

Refactoring and reengineering techniques are extensively documented in several
books (e.g. \cite{fowler}, \cite{feathers}, \cite{oorp}) and journal articles; the reader
is referred to them for detailed information.
The classical definitions of the two terms, as given in the above cited books,
is reported in table \ref{tab_def} for convenience.

A recurrent term in the context of refactoring, also mentioned in the following
sections, is ``smell''; it was introduced in \cite{fowler}.
Code smell is a surface indication that there might be a deeper problem in the
software system; by definition a smell is quick to spot (e.g. a long method).
Nevertheless code smells do not always indicate a real problem; usually they are
not bugs, i.e. they do not prevent the program from functioning correctly,
rather they indicate weaknesses in design that may hinder further
development or increase the risk of errors in the future.

\begin{center}
\begin{table}[h]
\label{tab_def}
\caption{Definition of relevant concepts.}
\centering
\begin{tabular}{l |l |c}
\hline
{\bf Concept} & {\bf Definition} 	& {\bf Source} \\
\hline
Refactoring 		& ``Refactoring is the process of changing a software system & \\
& in such a way that it does not alter the external behavior  & \cite{fowler} \\
& of the code yet improves its internal structure'' 	& \\
\hline
Reengineering	& Reengineering ``seeks to transform a legacy system into  & \\
& the system you would have built if you had the luxury    & \cite{oorp} \\
& of hindsight and could have known all the new requirements & \\
& that you know today''.	&	\\
\hline
\end{tabular}
\end{table}
\end{center}

\section{R\&D in Geant4 electromagnetic physics}
\label{sec_em}
Geant4 electromagnetic physics package has been subject to major evolution since
the first release of Geant4: it has included developments for new functionality,
and has been the playground for extensive design modifications.
Several ``code smells'' listed in Fowler's seminal ``Refactoring'' book can be
identified in this package (e.g. long methods, long parameter lists etc.);
standard refactoring techniques can be applied to attempt to improve the quality
of the software design.

Neither these symptoms nor the techniques to deal with them will be analyzed in
detail here; rather the attention is focused on less conventional ``smells''
identified in this code, which are specific to the physics simulation
environment, although conceptually similar to typical problems
addressed by refactoring techniques.

Duplicated code is  ``number one in the stink parade''  according to \cite{fowler}.
In the context of Geant4 electromagnetic physics in-depth analysis of the 
code and its physics performance has also identified duplicated physics
functionality and duplicated atomic parameters, which appear as 
different code.

\begin{figure}
\centerline{\includegraphics[angle=0,width=10cm]{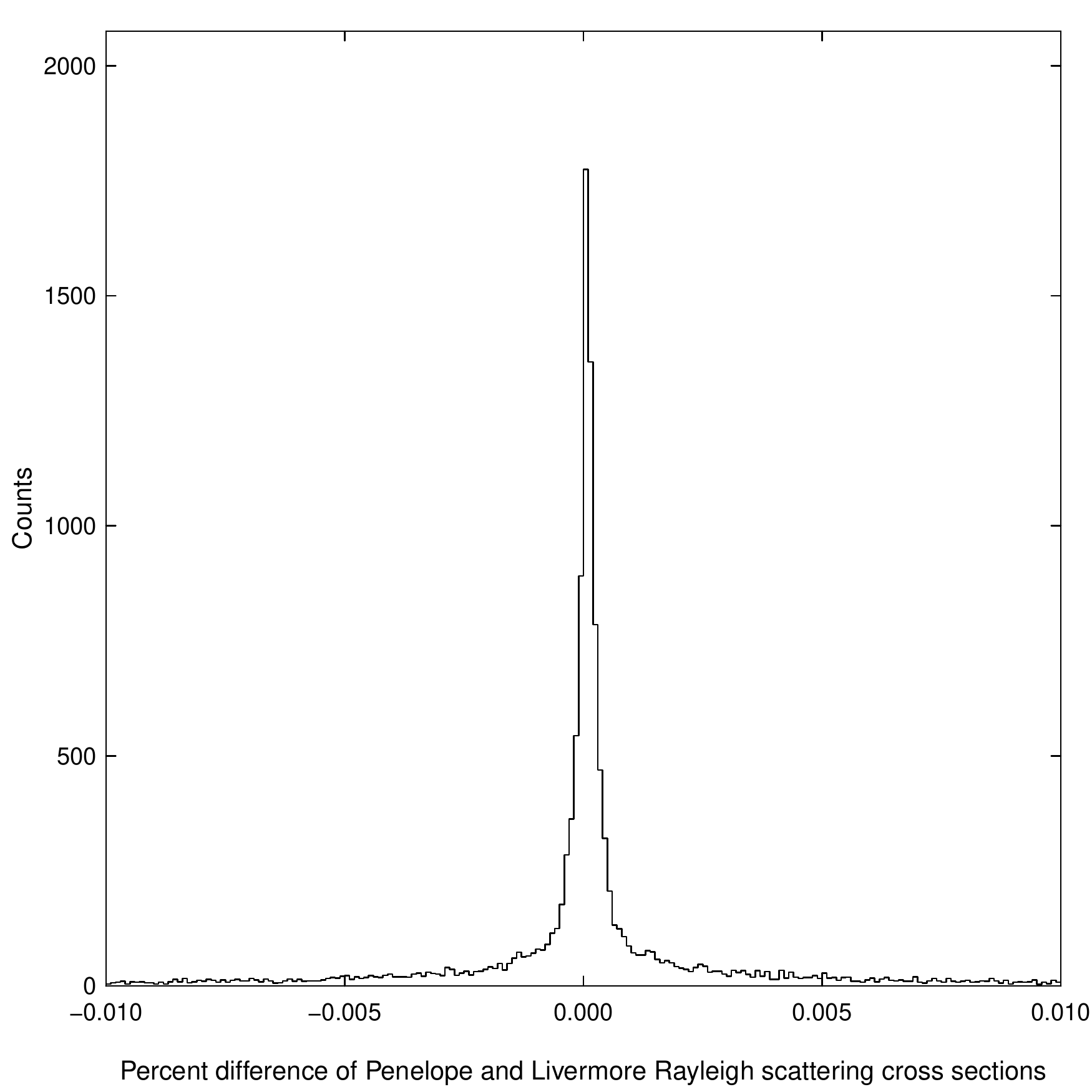}}
\caption{Percent difference of Rayleigh scattering cross sections calculated by
Geant4 ``Penelope'' and ``Livermore'' models.}
\label{fig_pentot}
\end{figure}

Duplicated physics models are models that provide identical functionality and
simulation accuracy in different implementations.
An example is the identical simulation of Rayleigh scattering in the Geant4
models identified as ``Livermore'' and ``Penelope'', respectively associated
with different classes: both models, as they are released in Geant4 9.5, are
based on the interpolation of cross sections and form factors tabulated in the
EPDL97 \cite{epdl97} data library.
Figure \ref{fig_pentot} shows the percent difference between the cross sections
calculated by the two models: the very small differences appearing in the
histogram are the results of EPDL97 data interpolation in the respective implementations.
The two implementations exhibit identical compatibility with experimental data,
quantified by means of goodness-of-fit-tests \cite{gof1,gof2}.
Further details on this issue can be found in \cite{tns_rayleigh}.
This duplication of functionality is the result of recent evolutions in the
Geant4 implementation of Penelope-like models: the Penelope-like Rayleigh
scattering implemented in the first reengineering of Penelope \cite{penelope}
was based on the original Penelope model, which did not use EPDL97; the
current Geant4 implementation was reengineered from a later version of Penelope (Penelope
2008), where the original Rayleigh scattering model had been replaced by one based on
EPDL97.

Duplicated physical parameters were also identified in Geant4 electromagnetic
package: for instance, different sets of atomic electron binding energies.
The duplication of atomic data in different parts of the code is prone to
generate inconsistencies in simulation observables depending on them.
Significant investment in software redesign is necessary to deal with 
atomic data consistently, while ensuring optimal accuracy for all the simulation
models that use them; 
further details are discussed in \cite{tns_binding}.

Automated techniques for the identification of duplicated code are available to
facilitate the refactoring process \cite{oorp}; nevertheless, to a large extent
the identification of ``bad smells'' in the code relies on the intuition of
experienced software developers.
The duplicated physics models and parameters in Geant4 electromagnetic 
package discussed above were identified thanks to rigorous physics validation 
analyses, complemented by in-depth code review.

Duplicated code and duplicated physics functionality implemented in Geant4
should be pruned.
Also code that exhibits inferior physics functionality than the model identified
as the state-of-the-art, and comparable or inferior computational performance
should be pruned.

An undesirable physics feature identified in the current design of Geant4
electromagnetic physics is the coupling between total cross section and final
state calculation in the same class.
Greater flexibility in choosing the two modeling approach independently would
ensure the optimization of physics configuration in experimental scenarios that
are especially concerned with simulation accuracy or computational performance.
It is worthwhile to note that the responsibilities for cross section calculation
and final state generation have been decoupled in Geant4 hadronic physics domain
since the RD44 phase.

Another undesirable feature of the current design of Geant4 electromagnetic
package is due to dependencies on other parts of the software: for instance, a
full scale Geant4-based simulation application, involving a geometry model, is
required even for testing low level physics modeling, such as a cross section calculation.

The last two features are associated with an inadequate problem domain analysis:
refactoring techniques are not sufficient to deal with these deficiencies, which
require improving the problem domain decomposition to provide sound foundation
for the software design.
A prototype design that addresses these issues, deriving from more effective
problem domain analysis, was presented at a previous conference
\cite{em_chep2009}; this design approach, which exploits generic programming
techniques, has been adopted in a recent large scale study of photon elastic 
scattering simulation \cite{tns_rayleigh}, where it demonstrated its ability to support the development
of a large variety of physics models and has enabled in-depth verification and validation
of their capabilities. 

A new model of photon elastic scattering based on S-matrix calculations (SM) has been developed
in the course of this study,
which improves the compatibility with experiment by approximately a factor two 
with respect to models currenty implemented in Geant4 (EPDL97), although at the price of some
deterioration of computational performance \cite{tns_rayleigh}.
Alternatively, more modern form factor calculations (MFASF) can improve the
compatibility with experimental data by approximately 40\% with respect to
current Geant4 models without additional computational burden.
The main results of the experimental validation process are summarized in table \ref{tab_eff};
the efficiency reported in the table represents the fraction of test cases where the $\chi^2$ test
finds a model compatible with experimental data with 0.01 significance.
The full set of results is documented in \cite{tns_rayleigh}.

\begin{table}[htbp]
  \centering
  \caption{Efficiency of  photon elastic scattering models at reproducing experimental data}
    \begin{tabular}{l|ccc}
    \hline
       Scattering angle & {\textbf{EPDL97}} 		& {\textbf{SM}}  		& {\textbf{MFASF}}  \\
    \hline
  	$0^{\circ} \leq \theta \leq180^{\circ}$		& 0.38 $\pm$0.06   	& 0.77 $\pm$0.05    	& 0.52 $\pm$0.06  \\
	$\theta\leq90^{\circ}$ 					& 0.40 $\pm$0.06   	& 0.82 $\pm$0.05    	& 0.54 $\pm$0.06  \\
	$\theta>90^{\circ}$ 					& 0.06 $\pm$0.06   	& 0.59 $\pm$0.05    	& 0.12 $\pm$0.06  \\
    \hline

    \end{tabular}
  \label{tab_eff}
\end{table}

\begin{figure}
\centerline{\includegraphics[angle=0,width=10cm]{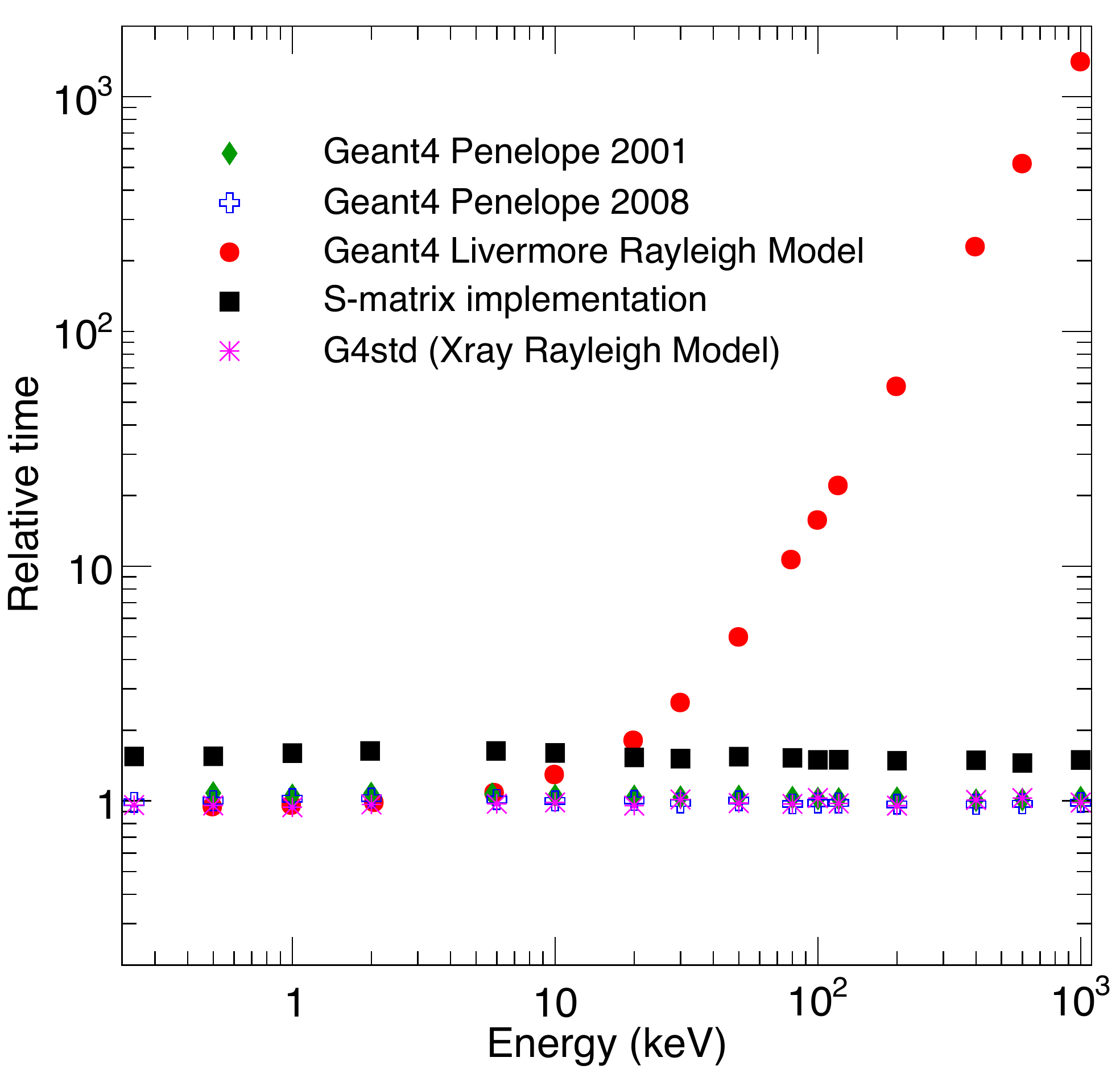}}
\caption{Computational performance of various Rayleigh scattering models``Livermore'' models.}
\label{fig_timeC}
\end{figure}

The refactoring process requires a sound and fine-grained testing system to
ensure that it does not modify the functionality of the code.
In the project in progress it is complemented by thorough testing for the
validation of Geant4 physics models and the evaluation of their
computational performance.
This fine-grained testing process allows the identification of the various
elements that contribute to the functionality of a class, and the quantification
of their accuracy and computational performance, to a great level of detail.
One of the results of this process is the quantification of the contributions to
computational performance intrinsically due to physics modeling, and those
related to other algorithms.
For instance, it has allowed the identification of an inefficient sampling
algorithm as the responsible for the apparently poor computational performance
of Rayleigh scattering implementation in the so-called Geant4 ``Livermore
model'' shown in figure \ref{fig_timeC}.
Once the inefficient sampling algorithm is replaced by a more efficient one,
the computational performance of that physics drops significantly, becoming equivalent 
to the performance of the fastest models shown in figure \ref{fig_timeC}.

A further attempt to improve computational performance in Geant4 physics domain 
by shifting the emphasis from algorithms to data libraries is currently the object od
exploration: it consists of minimizing the use of
algorithms in physics modeling, while privileging the use of data libraries.
A related study in progress evaluates the possibility of merging physics
models providing functionality for different energy rangess by smoothing data
tabulations derived from them, rather than through algorithms as it is currently
done in Geant4 \cite{chep2012_data}.
If the ongoing prototype investigations prove that these methods would achieve
significant performance improvements without degradation of physics accuracy,
a more extensive reengineering process will be justified.
Preliminary investigations of reengineering Geant4 data management domain 
\cite{chep2010_mincheol} have demonstrated significant gains in computational performance.

\section{R\&D in Geant4 radioactive decay simulation}

Significant effort has been invested into assessing the accuracy of Geant4
radioactive decay simulation, and improving its physics accuracy
and computational performance.

\begin{figure}
\centerline{\includegraphics[angle=0,width=16cm]{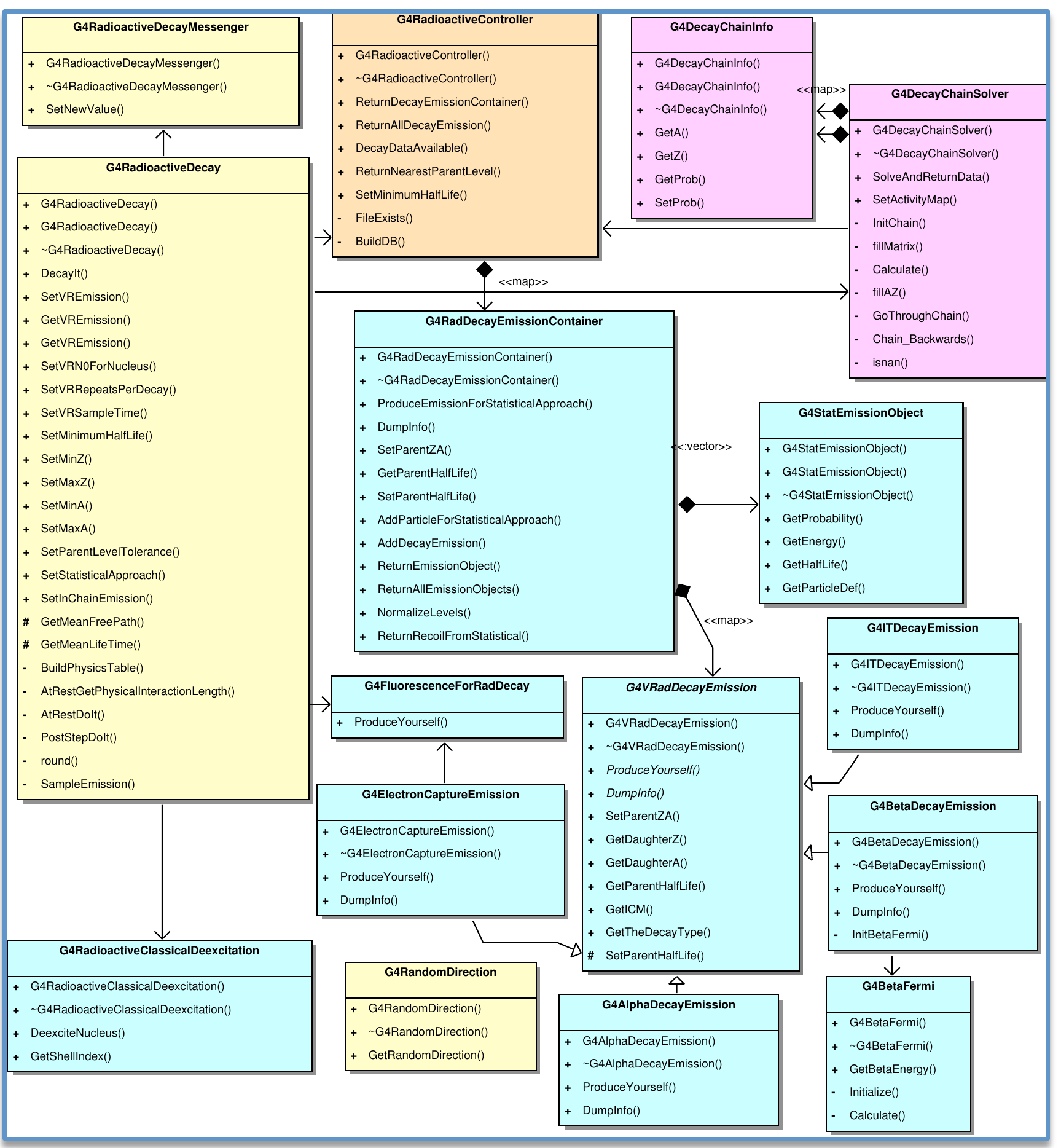}}
\caption{Class diagram of Geant4 radioactive decay resulting from the reengineering process.}
\label{fig_rdm_design}
\end{figure}

The reengineering process has improved the design, which is now based on a sound
domain decomposition and is characterized by well identified responsibilities.
A class diagram illustrating the main features of the reengineered software design
is shown in figure \ref{fig_rdm_design}.

\begin{figure}
\centerline{\includegraphics[angle=0,width=12cm]{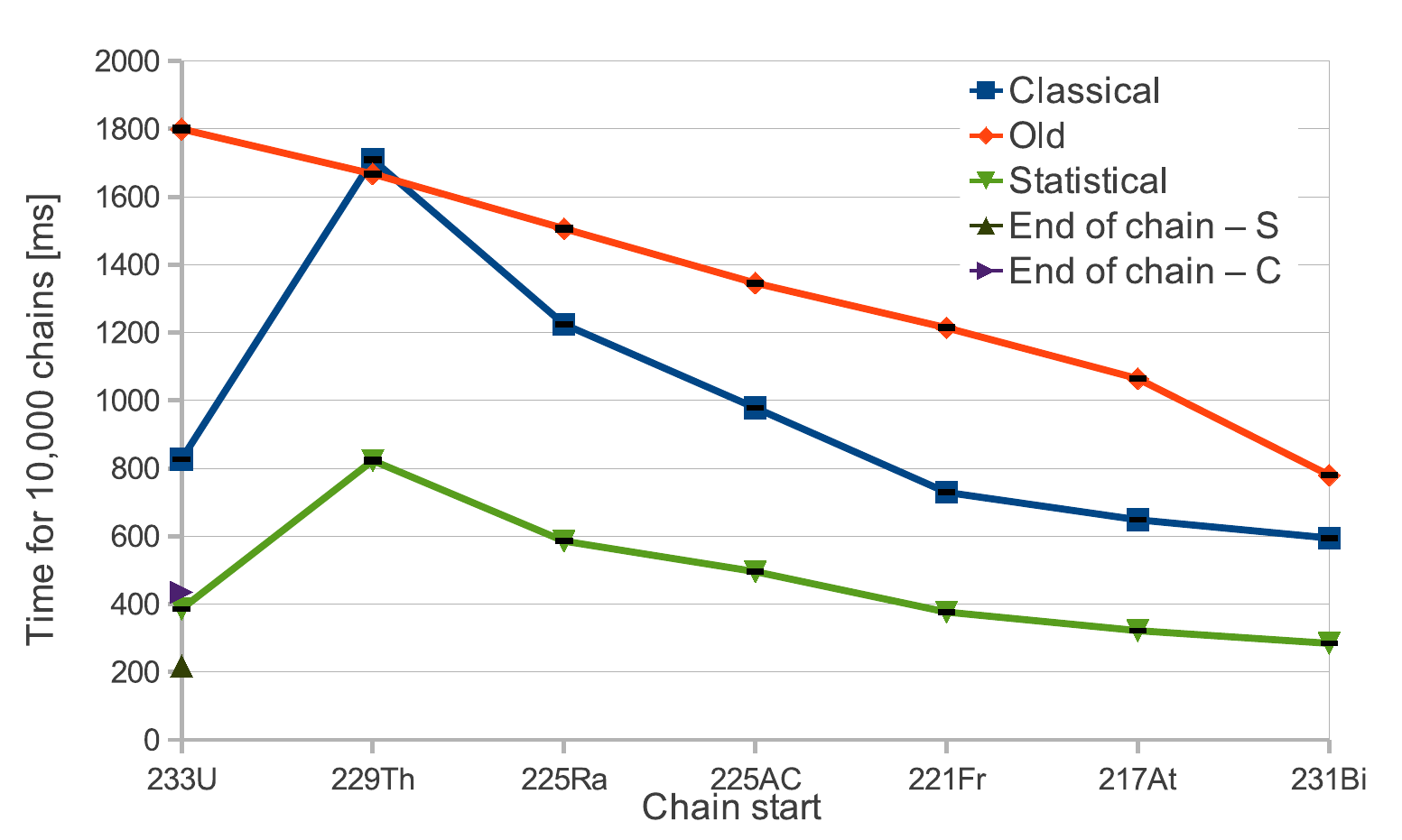}}
\caption{Computational performance of radioactive decay simulation when decaying $^{233}$U: current
Geant4 code (red), reengineered code (blue) and new algorithm (green).}
\label{fig_rdm_perf}
\end{figure}

Refactoring results in improved computational performance, as can be 
observed in figure \ref{fig_rdm_perf}.
Further improvement in computational performance are achieved by a new model,
which is based on a different conceptual approach with respect to the model
currently implemented in Geant4, since it treats the decay chain in statistical
terms.
The reengineered software design can accommodate both modeling alternatives,
whose different underlying approaches may address different use cases: this
extension of functionality would have not been possible in the context of the
original package design.

Extensive details of the reengineering process and its achievements are
documented in \cite{thesis_hauf}, along with the results of the experimental
validation of the software.

\begin{figure}
\centerline{\includegraphics[angle=0,width=16cm]{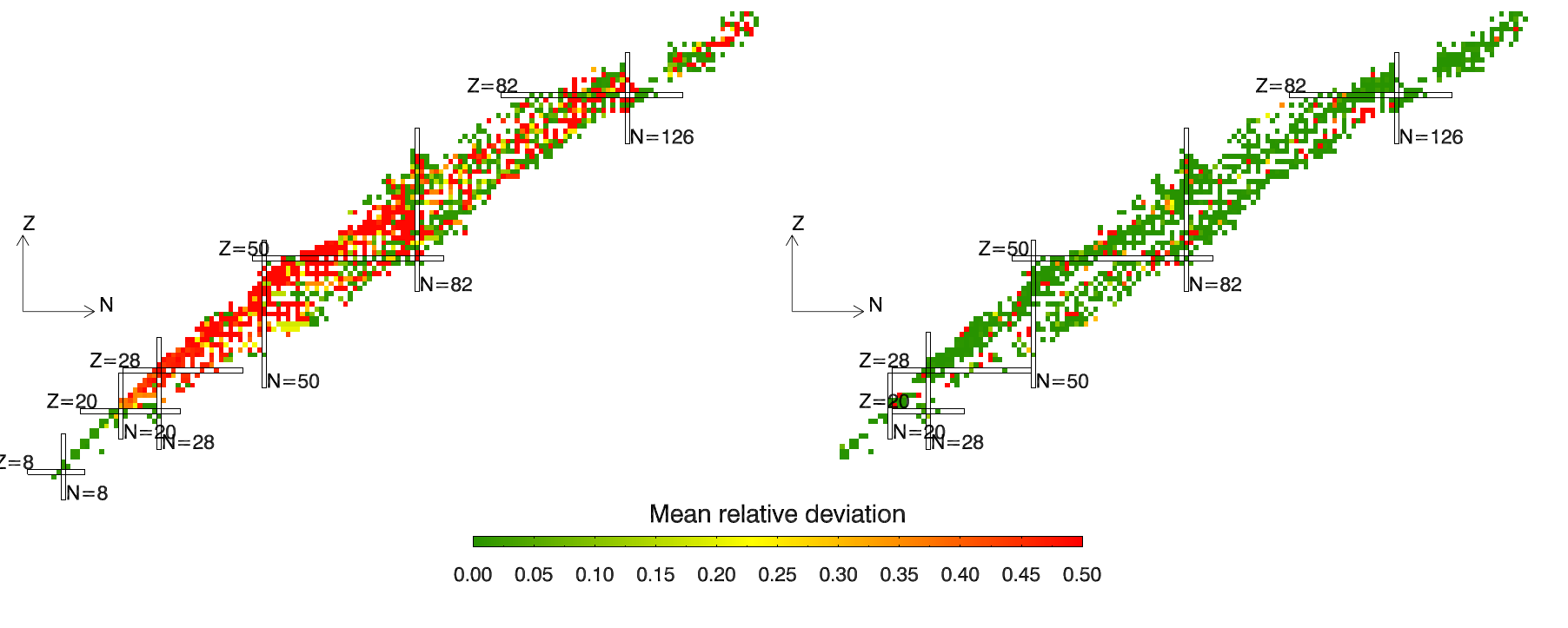}}
\caption{Nuclide charts showing the median relative intensity deviations from
reference data per isotope for Auger electron emission for the original Geant4
radioactive decay code (left), and for the reengineered one (right), which
exploits improved atomic parameters.}
\label{fig_rdm_auger}
\end{figure}

The reengineering of Geant4 radioactive decay code profits from the improvements
to atomic parameters discussed in section \ref{sec_em} regarding the refactoring
of Geant4 electromagnetic physics, resulting in better agreement with respect to
reference data: an example of results is illustrated in figure
\ref{fig_rdm_auger}, where nuclide charts show the median relative intensity
deviations from reference data per isotope for Auger electron emission for the
original Geant4 radioactive decay code, and for the reengineered one, which
exploits improved atomic parameters.

\section*{Conclusion}
Ongoing activities concerning the improvement of Geant4 design by means 
of refactoring and reengineering techniques have achieved significant results,
that demonstrate their contribution to improved physics accuracy and computational performance.

Further evaluations are in progress.

\section*{Acknowledgments}
The authors are grateful to the CERN Library for the support provided to this study.

\end{document}